\documentclass{iau} 
\usepackage{graphicx}

\title[Island Model \& IslandFAST] 
{The Neutral Islands during the Late Epoch of Reionization}

\author[Xu, Yue \& Chen]   
{Yidong Xu$^1$, Bin Yue$^1$
 \and Xuelei Chen$^{1,2,3}$}

\affiliation{$^1$Key Laboratory for Computational Astrophysics, National Astronomical Observatories, 
\\Chinese Academy of Sciences, Beijing 100012, China 
\\ email: {\tt xuyd@nao.cas.cn} \\[\affilskip]
$^2$University of Chinese Academy of Sciences, Beijing 100049, China \\
$^3$Center for High Energy Physics, Peking University, Beijing 100871, China}

\pubyear{2018}
\volume{333}  
\jname{Peering towards Cosmic Dawn}
\editors{Vibor Jeli\'c \& Thijs van der Hulst, eds.}
\begin{document}

\maketitle

\begin{abstract}
The large-scale structure of the ionization field during the epoch of reionization (EoR) can be modeled
by the excursion set theory. While the growth of ionized regions during the early stage
are described by the ``bubble model'', the shrinking process of neutral regions after
the percolation of the ionized region calls for an ``island model''.  
An excursion set based analytical model and a semi-numerical code ({\tt islandFAST}) have 
been developed. The ionizing background and the bubbles inside the islands
are also included in the treatment. With two kinds of absorbers of ionizing photons,
i.e. the large-scale under-dense neutral islands and the small-scale over-dense clumps, 
the ionizing background are self-consistently evolved in the model. 
\end{abstract}

\firstsection 
\section{Introduction}
Analytical modeling and approximated fast simulation are essential to explore
the large parameter space and complicated history of reionization.
As the large scale ionization field is closely correlated with the underlying 
density fluctuations (\cite[Battaglia et al. 2013]{2013ApJ...776...81B}), 
the excursion set theory(\cite[Bond et al. 1991]{1991ApJ...379..440B}; 
\cite[Lacey \& Cole 1993]{1993MNRAS.262..627L}) which give an intuitive 
description of the structure formation process are therefore also applied to 
the modeling of reionization process. In the ``bubble model''
(\cite[Furlanetto et al. 2004]{2004ApJ...613....1F}) the criterion of a region being ionized 
is translated to a density barrier, and the mass distribution of large ionized regions 
(``bubbles'') is obtained by computing the first up-crossing distribution of random trajectories
with respect to this bubble barrier.  Semi-numerical simulations of
the reionization process were also developed, which is much faster than 
radiative transfer simulations(e.g. \cite[Zahn et al. 2007]{2007ApJ...654...12Z}; 
\cite[Mesinger \& Furlanetto 2007]{2007ApJ...669..663M}; 
\cite[Mesinger et al. 2011]{2011MNRAS.411..955M}).

In the bubble model, the ionized regions during the early stage of EoR 
are assumed to be isolated and spherical. 
After the percolation of the HII regions, however, these basic assumptions are 
no longer valid, and the non-local source of ionizing photons must be accounted.  
In order to give a better description of the 
late stage of EoR, we proposed to consider the evolution of large-scale neutral regions (``islands'') 
which become isolated after percolation of ionized regions (\cite[Xu et al. 2014]{2014ApJ...781...97X}). 
The island model accounts for the ionizing background that is inevitable after 
percolation, and identifies neutral islands in the excursion set framework with first 
down-crossings of the island density barrier.
Based on the island model, a semi-numerical code, {\tt islandFAST}, is developed to 
make fast simulation of the evolution of the three-dimensional ionization field 
during the late EoR (\cite[Xu et al. 2017]{2017ApJ...844..117X}). 
It 
predicts the evolution of the ionization field as well as the ionizing background during the late stage of EoR. 
Here we present the basic idea and main results of the analytical and semi-numerical models.

\section{The Island Model}
\begin{figure*}[t]
\vspace*{-0.4 cm} \hspace*{-1.0 cm}
\centering{\includegraphics[scale=0.24]{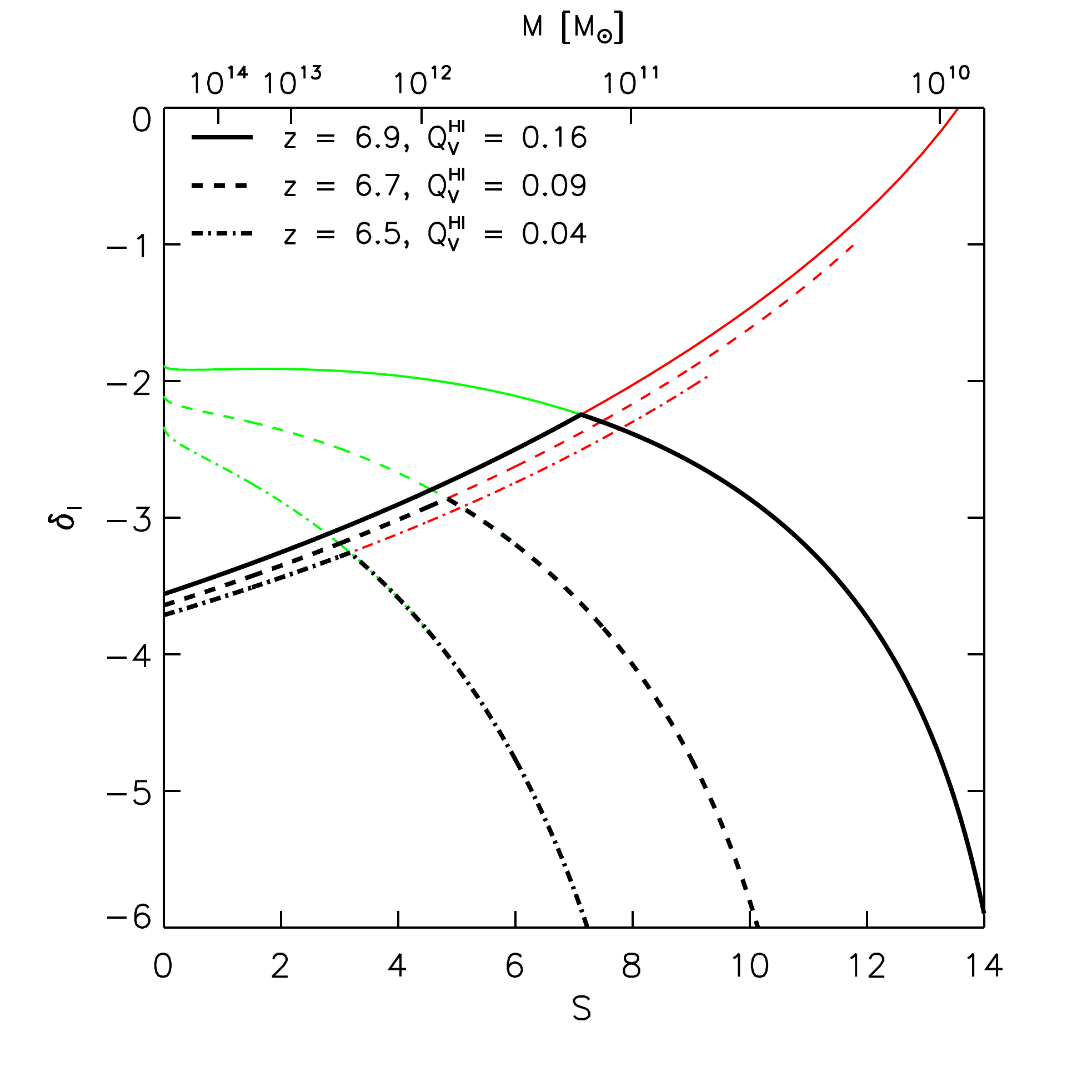}
\hspace*{-0.5 cm}
\includegraphics[scale=0.24]{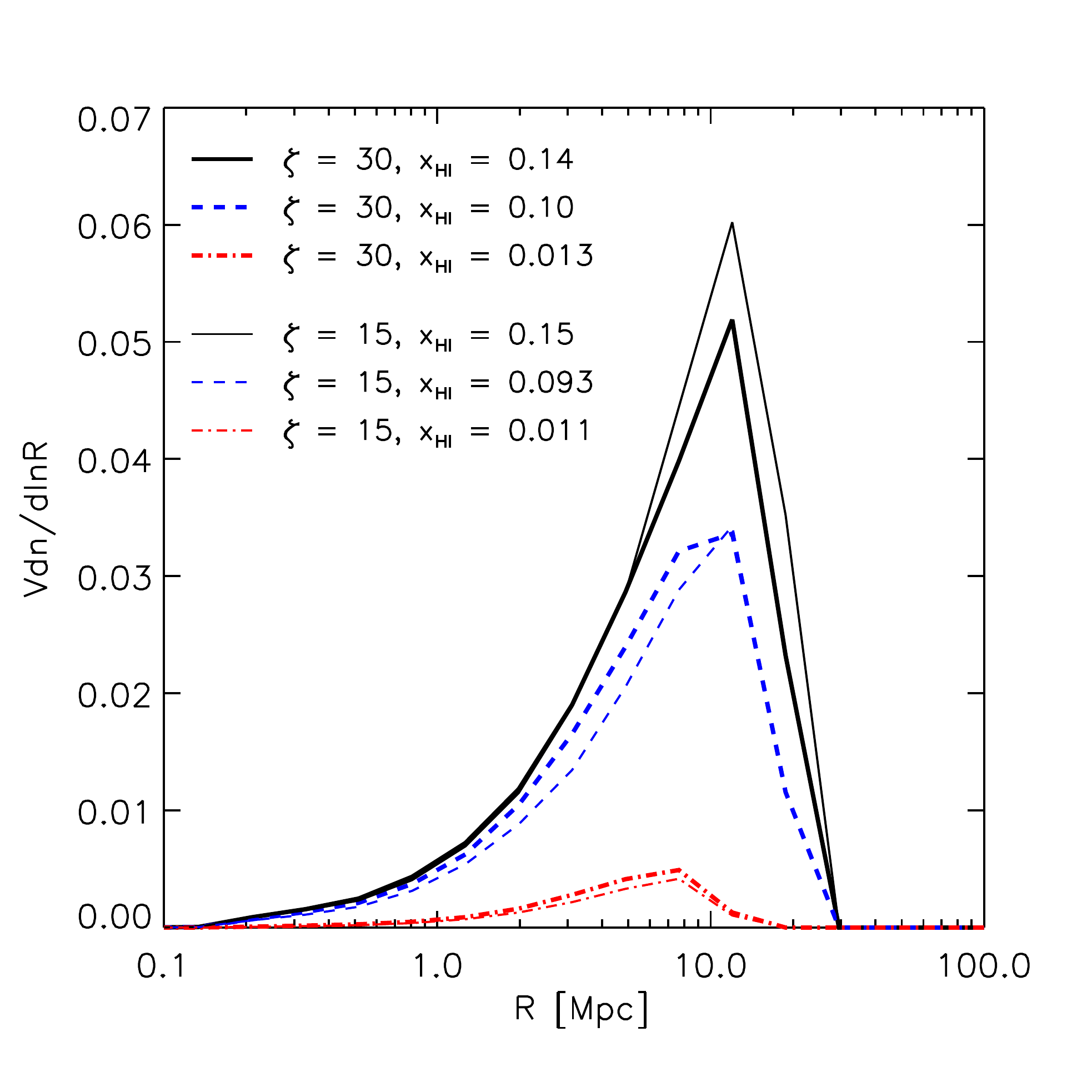}
\hspace*{-0.5 cm}
\includegraphics[scale=0.24]{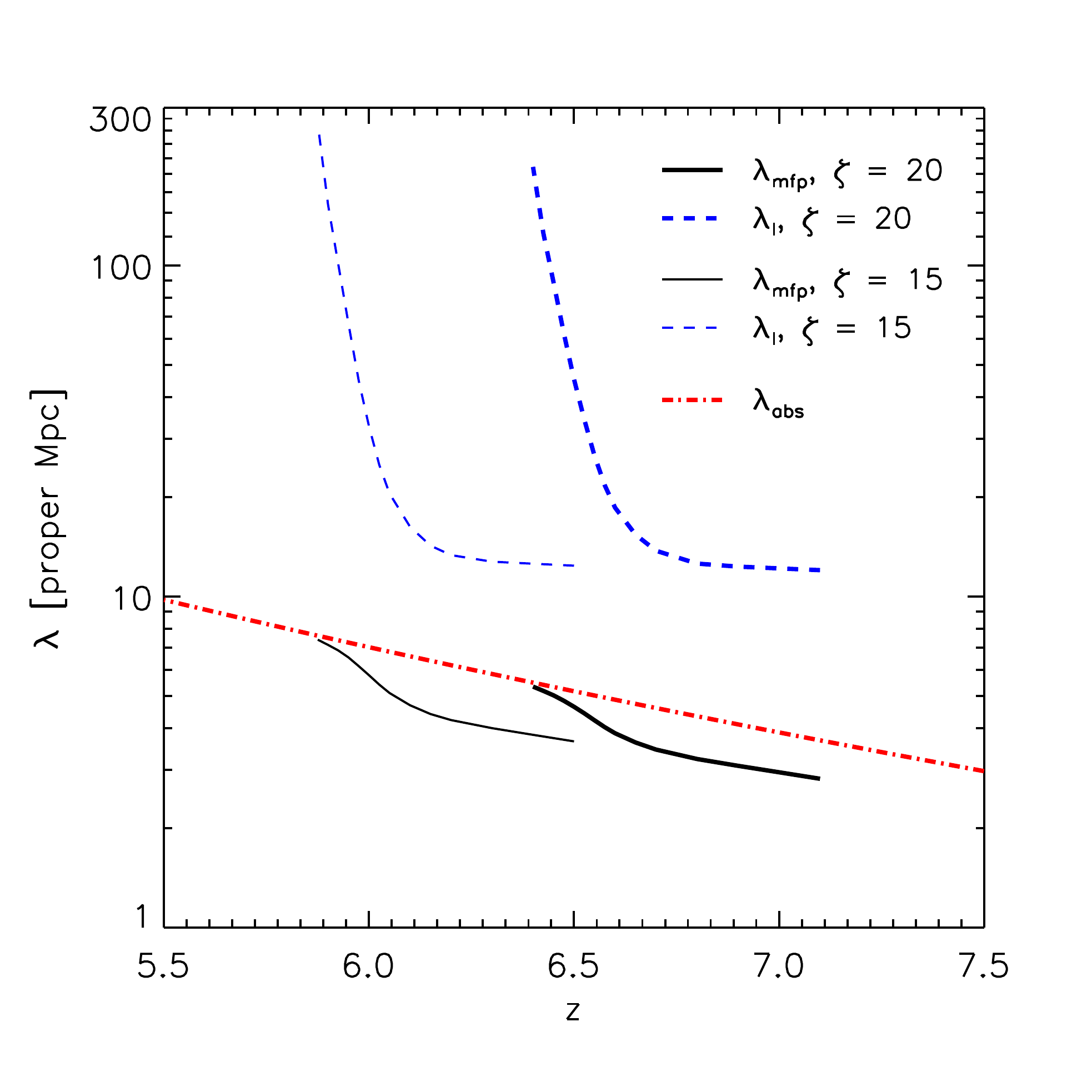}
\vspace*{-0.4 cm} \hspace*{-1.0 cm}
\caption{{\it Left panel:} The basic island barriers (green curves), the percolation threshold 
induced barriers (red curves), and the effective barriers (black curves).
The solid, dashed and dot-dashed curves are for $z=6.9, 6.7$ and 6.5 respectively.
{\it Central panel:} The size distribution of neutral islands in  the {\tt islandFAST} simulation, 
 with SAM neutral fraction threshold $f_{\rm HI}^c = 0.5$. 
The thick and thin lines are for $\zeta = 30$ and 15 respectively.
The solid, dashed, and dot-dashed curves correspond to mean neutral fractions $x_{\rm HI}$
as indicated in the legend.
{\it Right panel:} The evolution of the MFP of ionizing photons ({\it solid lines}).
The dashed lines show the MFP due to neutral islands, while
the dot-dashed line is the MFP due to small absorbers 
only. The thick lines are from the simulation with $\zeta = 20$,
and the thin lines are for $\zeta = 15$.}
\label{fig1}
}
\end{figure*}

In the island model, a region remains neutral if the available number of ionizing photons, 
either produced inside or outside, 
is not enough to ionize all hydrogen atoms in the region.
The condition for island formation consists of two terms: 
\begin{equation}\label{Eq.IslandCondition}
\xi f_{\rm coll}(\delta_{\rm M}; M,z)+ \frac{\Omega_m}{\Omega_b} 
\frac{N_{\rm back} m_{\rm H} }
{M X_{\rm H} (1+\bar{n}_{\rm rec})} < 1,
\end{equation}
where the first term on the L.H.S. accounts for the ionizing photons produced by 
collapsed objects inside, and the second term accounts for the contribution by the
ionizing background around it. 
$\zeta$ is the ionizing efficiency parameter,
$f_{\rm coll}$ is the collapsed fraction, and
$N_{\rm back}$ is the number of consumed background ionizing photons.
Assuming Gaussian density fluctuations,
this condition can be written as a
constraint on the density contrast of the region, i.e. the {\it island barrier}.
Assuming spherical islands, and that the consumption rate of background ionizing photons 
 is proportional to their surface area, 
we derive the basic island barrier shown with green curves in the left panel of Fig.\,\ref{fig1}. 
Different from the bubble barrier,
the island barrier bends down at small scales
because of the ionizing background.
One can derive the mass distribution and the volume fraction of neutral regions
by solving for the first-{\it down}-crossing 
distribution of random trajectories with respect to the island barrier.

Note that there is a ``bubbles-in-island'' effect,
as there might be self-ionized regions inside a host island. 
The distribution of bubbles inside islands are obtained
by computing the conditional probability distribution of the trajectories.
The average bubbles-in-island fraction can be derived by integrating over
all possible bubble sizes for a given island, and this fraction is limited by
a percolation threshold $p_c$ in order to define the bona fide islands,
in the sense that an island is not percolated through by bubbles inside.
This percolation criterion of $q_{\rm B} < p_c$ acts 
as an additional barrier for finding islands (plotted with the red curves in 
the left panel of Fig.\,\ref{fig1}), which combines with the basic island barrier
to define the host islands.
The percolation threshold also demarcates the scope of application of 
both bubble model and island model.
The bubble model can be reliably applied before bubble filling factor becomes larger than 
 $p_c$, while the island model should be used below the redshift at which the
island filling factor equals to $p_c$.

With the combined island barrier, 
the island model predicts the size distribution and evolution of islands
after the islands become isolated.
As the redshift decreases, small islands disappear rapidly 
while the large ones shrinks.
Eventually, all these under-dense islands are ionized, 
only compact neutral regions such as galaxies or minihalos remain.

\section{The IslandFAST Model}
The semi-numerical simulation {\tt islandFAST} is developed from the  
{\tt 21cmFAST} (\cite[Mesinger et al. 2011]{2011MNRAS.411..955M}), 
but extended to treat the late EoR by the island 
model  (see \cite[Xu et al. (2017)]{2017ApJ...844..117X} for details).
Compared with the {\tt 21cmFAST}, 
the major differences of the {\tt islandFAST} are 
the two-step filtering algorithm in generating the ionization field in order to 
take bubbles-in-island effect into account, and the self-consistent treatment for
the ionizing background accounting for various absorption systems.
The mean free path of ionizing photons and the corresponding intensity of the 
ionizing background is obtained by an iterative procedure.
Adaptive redshift steps are applied in order to achieve convergence in 
the ionization field and the intensity of the ionizing background.

\begin{figure}[t]
\vspace*{-0.4 cm} \hspace*{-1.5 cm}
\centering{\includegraphics[scale=0.37]{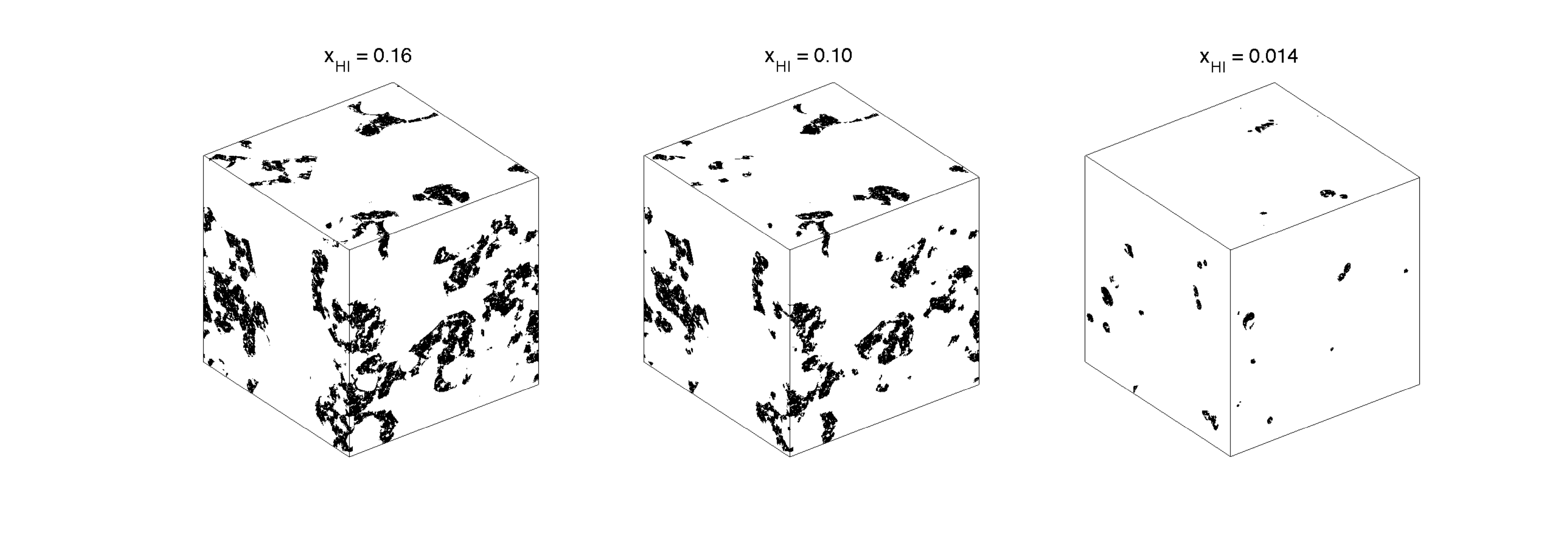}
\vspace*{-1.5 cm} \hspace*{-1.0 cm}
\caption{The 3-dimensional visualization of the ionization fields from the {\tt islandFAST} with 
a box size of $200\, {\rm h}^{-1} {\rm Mpc}$, $512^3$ resolution, and $\zeta = 20$. 
The neutral islands are shown as black patches, and the ionized regions are left white.
The three boxes have the mean neutral fractions of 
 0.16, 0.10, and 0.014, from left to right respectively.}
\label{fig2}
}
\end{figure}

We find that the late reionization proceeds very fast, and
the bubbles-in-island effect is quite obvious, showing
complex morphology of the ionization field (Fig.~\ref{fig2}).
In the central panel of Fig.~\ref{fig1} we show the comoving size distributions 
of neutral islands for various global neutral fraction. 
We find a characteristic scale of $\sim 10$ Mpc for the
neutral islands at each redshift. 
This scale decreases very slowly as the islands being ionized.
Judging from the simulation box, this is perhaps because
the neutral islands have very complex shape, and using the spherical average method (SAM) to
identify islands tend to divide a large island into pieces. As the reionization proceeds, 
the larger islands gradually break into smaller ones that would be identified as similar sizes
by the SAM. Also, the large islands shrink gradually, 
and partially compensates for the disappearance of the smaller islands.

While generating the ionization field, the {\tt islandFAST} simultaneously predicts 
the level of the ionizing background. 
We find that a higher ionizing efficiency would lead to earlier growth and 
higher intensity  of the ionizing background. 
The intensity of the ionizing background is boosted by an order of magnitude 
if we neglect small absorbers, resulting in a much faster completion of reionization. 
Similar trends are seen in the evolution of the mean free path (MFP) of the ionizing 
photons as shown in the right panel of Fig.~\ref{fig1}. 
The MFP  due to the under-dense islands alone,  
$\lambda_{\rm I}$, and that limited by over-dense absorbers, $\lambda_{\rm abs}$, are also shown.
We find that the MFP are mainly limited by the small-scale 
absorbers, and the shading effect of the large-scale islands only
reduce the MFP moderately during the late EoR. 
Therefore, the level of the ionizing 
background is primarily regulated by small-scale absorbers, and they significantly delay
and prolong the reionization process.

\section{Summary}

We have developed the island model--an analytical model of reionization using the excursion set approach, 
and the corresponding semi-numerical simulation, {\tt islandFAST},
for the evolution of isolated neutral regions after the percolation of ionized bubbles,
complementary to the bubble model for the early reionization epoch. 
As the case of bubble model, 
although the excursion set model uses spherical window function for computing the mean density at different 
scales, the model can and does produce complicated reionization topology. 
Our model incorporated the effect of ionizing background photons and  
accounted for the bubbles-in-island effect. A percolation
threshold is introduced to demarcate the scope of application for the island model 
and the bubble model, and also to define the bona fide islands. 
Given a model for the small-scale absorbers, the {\tt islandFAST} 
generates the ionization field after percolation, and simultaneously predicts the intensity 
of the ionizing background.

The late reionization proceeds quite fast. As the large islands shrink with time, 
 the small ones are swamped by the ionizing photons. If characterized by the SAM
with a high neutral fraction threshold, the islands exhibits 
a relatively robust characteristic scale of $\sim 10$ Mpc throughout the late EoR. 
We find that the large-scale islands dominate the morphology of the ionization field,
while the small scale absorbers dominate the opacity of the IGM, and play a 
major rule in determining the level of the ionizing background and the speed of reionization 
process. 

With the advent of next generation EoR experiment, reionization models will be put to 
rigorous  tests, and the late state is of particular interest as it might be observationally more accessible. 
We expect these models would be helpful for interpreting such observations.

{\bf Acknowledgements}
YX is supported by the NSFC grant 11303034, and the Young Researcher Grant of 
National Astronomical Observatories, CAS. 
BY is supported by the NSFC grant 11653003 and CAS Pioneer Hundred Talents (Young Talents) Program. 
XC is supported by the MoST grant  2016YFE0100300 and 2012AA121701,
the NSFC key project grant 11633004 and grant 11373030, the CAS
Frontier Science Key Project QYZDJ-SSW-SLH017,  and 
Strategic Priority Research Program XDB09020301. 
We thank our collaborators Meng Su and Zuhui Fan with whom the first part of the work 
was done. We thank the organizers of the IAUS 333 for a successful and productive conference.

\end{document}